\input phyzzx
\hsize=40pc


\catcode`\@=11 
\def\NEWrefmark#1{\step@ver{{\;#1}}}
\catcode`\@=12 

\def\square{\kern1pt\vbox{\hrule height 1.2pt\hbox{\vrule width 1.2pt\hskip 3pt
   \vbox{\vskip 6pt}\hskip 3pt\vrule width 0.6pt}\hrule height 0.6pt}\kern1pt}
\def\inbar{\,\vrule height 1.5ex width .4pt depth0pt}\def\IC{\relax
\hbox{$\inbar\kern-.3em{\rm C}$}}
\def\IH{\relax\hbox{$\inbar\kern-.3em{\rm H}$}}

\def\bra#1{\langle #1 |}
\def\ket#1{| #1 \rangle}

\def\A{{\cal A}}
\def\B{{\cal B}}
\def\C{{\cal C}}
\def\D{{\cal D}}

\def\I{{\cal I}}

\def\K{{\cal K}}

\def\O{{\cal O}}

\def\T{{\cal T}}

\def\V{{\cal V}}

\def\p{\partial}

\def\wt{\widetilde}
\def\wh{\widehat}

\def\B{{\cal B}}

\def\V{{\cal V}}
\def\O{{\cal O}}

\def\p{\partial}

\singlespace

\def\define#1#2\par{\def#1{\Ref#1{#2}\edef#1{\noexpand\refmark{#1}}}}
\def\con#1#2\noc{\let\?=\Ref\let\<=\refmark\let\Ref=\REFS
         \let\refmark=\undefined#1\let\Ref=\REFSCON#2
         \let\Ref=\?\let\refmark=\<\refsend}

\let\refmark=\NEWrefmark

\define\cbi{A. Sen and B. Zwiebach, `Local background independence of
classical closed string field theory',
Nucl. Phys. {\bf B414} (1994) 649, hep-th/9307088.}

\define\qbi{A. Sen and B. Zwiebach, `Quantum background independence of closed
string field theory',
Nucl. Phys. {\bf B423} (1994) 580, hep-th/9311009.}

\define\nonconf{B. Zwiebach, `Building string field theory around
non-conformal backgrounds',
Nucl. Phys. {\bf B480} (1996) 541, hep-th/9606153.}

\define\moduli{B. Zwiebach, `New moduli spaces from string background
independence consistency conditions',
Nucl. Phys. {\bf B480} (1996) 507, hep-th/9605075.}

\define\ranga{K. Ranganathan, `Nearby CFT's in the operator formalism: the
role of a connection',
Nucl. Phys. {\bf B408} (1993) 180, hep-th/9210090.}

\define\rsz{K. Ranganathan, H. Sonoda and B. Zwiebach, `Connections
on the state-space over conformal field theories',
Nucl. Phys. {\bf B414} (1994) 405, hep-th/9304053.}

\define\dilaton{O. Bergman and B. Zwiebach `The dilaton theorem and closed
string backgrounds',
Nucl. Phys. {\bf B441} (1995) 76, hep-th/9411047.}

\define\geom{S. Rahman, `Geometrising the closed string field action',
{\it work in progress.}}

\singlespace
{}~ \hfill \vbox{\hbox{MIT-CTP-2626}
\hbox{
} }\break
\title{CONSISTENCY OF QUANTUM BACKGROUND INDEPENDENCE}
\author{Sabbir A Rahman \foot{E-mail address: rahman@marie.mit.edu
\hfill\break Supported in part by D.O.E.
cooperative agreement DE-FC02-94ER40818.}}
\address{Center for Theoretical Physics,\break
Laboratory of Nuclear Science\break
and Department of Physics\break
Massachusetts Institute of Technology\break
Cambridge, Massachusetts 02139, U.S.A.}

\abstract
{We analyse higher order background independence conditions arising from
multiple commutators of background deformations in quantum closed string field
theory. The conditions are shown to amount to a vanishing theorem for
$\Delta_S$ cohomology classes. This holds by virtue of the existence of moduli
spaces of higher genus surfaces with two kinds of punctures. Our result is a
generalisation of a previous genus zero analysis relevant to the classical
theory.}
\endpage

\singlespace
\baselineskip=18pt

\chapter{\bf Introduction and Summary}

It has been some time now since the background independence of string field
was proven [\cbi,\qbi]. The infinitesimal background deformations were shown
to be implemented as canonical transformations whose Hamiltonian functions 
were defined by moduli spaces of punctured Riemann surfaces with a single
special puncture. It was realised in [\moduli] that these $\B^1$ spaces
represented only the first order perturbations of the string background,
and that higher order deformations implied the existence of moduli spaces
$\B^2$, $\B^3$,... of surfaces having more than one special puncture,
antisymmetric under the exchange of special punctures. These were
duly constructed in [\moduli, \nonconf] for the classical case. Furthermore,
these $\B$-spaces have precisely the properties required to formulate
string theory around non-conformal backgrounds, and such an action was
constructed explicitly in [\nonconf]. The $\B$-spaces do not depend upon any
particular
choice of string background and as such would be expected to play an
important role in any manifestly background independent formulation of the
theory. Arriving at such a formulation remains one of the major goals in
string theory, and provides motivation for the present work.

The recent work of Zwiebach [\moduli, \nonconf] dealt only with the
classical closed string theory, whereas we would of course eventually wish
to deal with the full quantum theory. Experience has shown that in string
field theory, classical results usually generalise to the quantum
case without too many complications, and this paper reaffirms this by
finding the quantum generalisations of Zwiebach's results.

This work is organised as follows. In \S 2 we review some properties of
connections on the space of conformal field theories, and recall some basic
facts about the string vertices, introducing some new notation for $\B$-spaces
which will be used throughout the paper. In \S 3, we follow [\qbi] to derive
the conditions for background independence when the Batalin-Vilkovisky
density function $\rho$ is allowed some field-dependence. We then review
the origin of general symplectic connections, and give the form of the
Hamiltonian $B_\mu(\Gamma)$ implementing background deformations for general
connections. We find that background independence to first order amounts to
a $\Delta_S$-cohomology theorem for the action, being the natural
generalisation of the antibracket-cohomology of the classical
case. In \S 4 we
consider the commutator of background deformations and show that background
independence to second order implies a higher cohomology theorem.
The consistency conditions are shown to be satisfied through the existence of
spaces $\B^2$ with two special punctures of all genera. In \S 5, we use
the language of differential forms on the theory space manifold to efficiently
derive $\Delta_S$-cohomology theorems to all higher orders for the string
action, and extend the complex of $\B$-spaces to
include positive-dimensional moduli spaces of Riemann surfaces for all
genera and all numbers of ordinary and special punctures compatible with
the dimensionality requirement.

\chapter{\bf Review and Notation}

\noindent
In this section we will review a few definitions and results used 
in the present work.

\section{Connections on the Space of Conformal Theories} 

We review here the formalism of [\ranga, \rsz]. A vector bundle is constructed
over the space $M$ of CFTs by assigning a
basis $\ket{\Phi_i^x}$ of states of the theory to each point $x = \{ x^\mu \} =
(x^1, . \, . \, . \, , x^m)$ of $M$.
The coordinates of the vector space at $x^\mu$ are denoted by $\psi = \{
\psi^i \} = (\psi^1, . \, . \, . \, , \psi^{2n})$.
Given a connection $\Gamma_{\mu i}^j(x)$ on this bundle, the covariant
derivative of sections, $\bra{A(x)} = \sum_i a_i(x) \bra{\Phi^i_x}$ and
$\ket{B(x)} = \sum_i \ket{\Phi^x_i} b^i(x)$, on the bundle are defined by,
$$\eqalign{D_\mu (\Gamma) \bra{A } &\equiv \p_\mu a_i \bra{\Phi^i} - 
a_i \Gamma_{\mu j}^i \bra{\Phi^j}\cr
&=  \p_\mu \bra{A} - \bra{A} \, \Gamma_\mu\,,}\eqn\sabzeroa$$
\noindent
where $\Gamma_\mu = \sum_i \ket{\Phi_i} \, \Gamma^i_{\mu j} \bra{\Phi^j}$, and,
$$\eqalign{D_\mu (\Gamma) \, \ket{B} &\equiv \ket{\Phi_i} \, \p_\mu b^i +
\ket{\Phi_j} \, \Gamma_{\mu i}^j \, b^i \cr
&= \p_\mu \, \ket{B} + \Gamma_\mu \, \ket{B}\,.}\eqn\sabzerob$$
The covariant derivatives of functions on the bundle are defined by,
$$\eqalign{D_\mu (\Gamma) F &\equiv \p_\mu F - F \overleftarrow{\p_i}
\Gamma_{\mu j}^i \, \psi^j \cr
&= \partial_\mu F - F {\overleftarrow\p \over \p \, \ket{\Psi}} \,
 \Gamma_\mu \, \ket{\Psi}\,,}\eqn\sabzeroc$$
where we use Greek indices for theory space coordinates and Latin indices for
symplectic coordinates and we introduce the notation,
$${\overleftarrow\p\over\p \, \ket{\Psi}} \equiv \overleftarrow{\p_i}
\bra{\Phi^i}\,.\eqn\sabzeroca$$
We will primarily be interested in symplectic connections for which the
covariant derivatives of the symplectic form and the sewing ket vanish,
$$D_\mu (\Gamma) \, \bra{\omega_{12}} = D_\mu (\Gamma) \, \ket{S_{12}} =
0\,.\eqn\sabzerod$$
In particular this implies that the covariant derivative of a function of
the type
$\bra{A}\Psi\rangle \cdots \ket{\Psi}$ defined by tensor sections $\bra{A(x)}$
is given by,
$$D_\mu \bra{A}\Psi\rangle\cdots \ket{\Psi} = 
(D_\mu \bra{A} ) \, \ket{\Psi}\cdots \ket{\Psi}\,,\eqn\sabzeroe$$
where covariant derivatives of sections are simply given by,
$$D_\mu (\Gamma) \bra{A} = \partial_\mu \bra{A} - 
\sum_n \bra{A} \, \Gamma_\mu^{(n)}\,,\eqn\sabzerof$$
with the label $n$ referring to a state space in the tensor section.

\section{Moduli Spaces of String Vertices and $\B$ Spaces}
The string vertices $\V = \sum \hbar^g \, \V_{g,n}$ (where the sum is over
$n\geq3$ for genus zero, $n\geq1$ for genus one, and $n\geq0$ for all
higher genera) satisfy the recursion relations (see for example Eqn.(2.22) of
[\qbi]),
$$\partial \, \V + \hbar \Delta \V + \half \{ \V , \V \} = 0\,.\eqn\sabzerog$$
The string action may then be written (Eqn.(3.35) of [\qbi]),
$$S = Q + \hbar S_{1,0} + f(\V)\,, \eqn\sabzeroh$$
which satisfies the B-V master equation,
$$\half \{ S , S \} + \hbar \Delta S = 0\,.\eqn\sabzeroi$$
The interpolating `$\B$-spaces' have been introduced (using various
notations) in the papers [\cbi,\qbi,\dilaton,\moduli,\nonconf], and are
responsible for first and higher order infinitesimal background deformations
of the theory via the canonical transformations implemented by their associated
Hamiltonians. They also play an important r\^ole in the
formulation of string field theory around non-conformal backgrounds.

We will define $\B^{\bar n}_{g,n}$ as the
moduli space of decorated Riemann surfaces of genus $g$ with $n$ ordinary
punctures (each surrounded by a coordinate disk) and with $\bar n$
special punctures. The $\B$-spaces are
symmetrised with respect to labellings of the ordinary punctures and
antisymmetrised with respect to labellings of the special punctures, and we
recall that the space $\B^{\bar n}_{g,n}$ has dimension $6g-6+2n+3{\bar n}$.
The spaces $\B^0_{g,n}$ with no special punctures will be identified with
the usual string vertices $\V_{g,n}$. In the papers
[\cbi,\qbi,\dilaton] the $\B$-spaces having a single special puncture
with $n\geq2$ at genus zero and $n\geq1$ at higher genus were introduced,
being responsible for implementing first order background deformations. In
[\moduli,\nonconf], higher order deformations of the classical theory
were considered and it was found necessary to extend this complex
to include the spaces $\B^{\bar n}_{0,n}$ for all $n\geq1$ and $\bar n\geq2$
which implemented them.

Like the string vertices $\V$, the $\B$-spaces also satisfy recursion
relations, and the ones which will be needed for the purposes of this paper
are those satisfied by the spaces $\B^1$ (see for example Eqn.(3.20) of
[\dilaton]),
$$\delta_\V \B^1 = (\K - \I) \V + \V'_{0,3} + \hbar \Delta \B^1_{0,2} +
\hbar \I \V_{1,1}\,,\eqn\sabzeroia$$
where $\B^1 \equiv \sum_{g,n} \B^1_{g,n}$ and $\delta_\V \equiv \p + \hbar
\Delta + \{ \V , \, \cdot \, \}$. There is no space $\B^1_{1,0}$ so the terms
involving objects of genus one and with one special puncture
but no ordinary punctures have been extracted by hand to leave an equation
which holds for all $(g,n)$.

Other useful formulae relating the various operators acting on moduli
spaces and functions are collected in the Appendix for reference.

\chapter{\bf Background Independence Revisited}

In this section, we review the derivation of the condition for quantum
background independence [\qbi]. This was originally carried out for the case
where the B-V density $\rho$, being dependent only on $x$, was a section on
the theory space bundle. Here we shall derive the background independence
conditions for the more general case where $\rho=\rho(\psi,x)$ is allowed
some field-dependence and hence is promoted to a function on the bundle.
Given that the invariant B-V measure takes the form $d \mu_S \equiv \prod_i
d \psi^i \, \rho \, e^{2S/\hbar}$, there should be no problem in transferring
some of the field-dependence of the action to the B-V density, and we verify
this by analysing the transformation properties of our background independence
condition under field-redefinitions. Nevertheless, since it is always possible
to choose a frame in which $\rho$ is field-independent, we choose
for simplicity to restrict our analysis to this case. After briefly reviewing
the origin of general symplectic connections [\qbi], we write the
explicit form of the Hamiltonian $B_\mu(\Gamma)$ implementing background
deformations for general symplectic connections, and show that background
independence to first order amounts to a `vanishing' theorem for
$\Delta_S$ cohomology classes.

\section{Field-Independence of the Density $\rho$}

Having chosen field/antifield coordinates $( \psi^i )$ on the
symplectic manifold we can define the string field $\ket{\Psi}$ as usual by,
$$\ket{\Psi} = \sum_i \ket{\Phi^x_i} \psi^i_x\,.\eqn\sabhalfa$$
where the superscript $x$ denotes the theory space dependence. We will make $x$
implicit in the following. Let us investigate the effect of allowing the
density $\rho = \rho(\psi,x)$ to
have an explicit dependence on the field/antifield coordinates (though we are
always able to choose coordinates such that the field-dependence vanishes).
In the field-independent case, $\rho$ drops out of the expression for the
delta operation, but this explicit $\rho$-dependence is reinstated on allowing
the field-dependence, which may be determined as follows,
$$\eqalign{\Delta F &= {1\over2 \rho} (-)^i \overrightarrow{\p_i}
(\rho \, \omega^{ij} \overrightarrow{\p_j} F)\cr
&= \half (-)^i \left((\overrightarrow{\p_i} \, \hbox{ln} \, \rho) \,
\omega^{ij} \overrightarrow{\p_j} F + \overrightarrow{\p_i} (\omega^{ij}
\overrightarrow{\p_j} F) \right)\cr
&= \{ \half \, \hbox{ln} \, \rho , F \} + \wh\Delta F\,,}\eqn\sabhalfc$$
where we have defined the $\rho$-independent hatted delta operation by,
$$\wh
\Delta F \equiv  \half (-)^{F+1}
\left({\overrightarrow\p\over\p \ket{\Psi}_1} {\overrightarrow\p F\over\p
\ket{\Psi}_2} \right) \ket{S_{12}}\,,\eqn\sabhalfca$$
and have used the notation,
$${\overrightarrow\p\over\p \ket{\Psi}} \equiv \bra{\Phi^i}
\overrightarrow{\p_i}\,.\eqn\sabhalfd$$
We note that $\wh\Delta$, unlike $\Delta$, is in general neither a nilpotent
nor a scalar operator.

We may now reconsider the consistency condition for local quantum
background independence. This condition (Eqn.(4.16) of [\qbi]) was derived for
the case of the canonical connection $\wh\Gamma_\mu$ of [\ranga], and reads,
$$D_\mu (\wh\Gamma) S = \hbar \Delta B_\mu + \{ S , B_\mu \} + \half \hbar
\bigl( \hbox{str} \, \wh\Gamma_{\mu} - \p_\mu \, \hbox{ln} \, \rho \bigr)
\,.\eqn\sabhalfe$$
We would like to derive the general form where $\rho = \rho(\psi,x)$ has
a field/antifield coordinate dependence. Given the measures $d \mu_x
= \rho (x, \psi_x) \prod_i d \psi_x^i$ and $d \mu_y = \rho (y, \psi_y)
\prod_i d \psi_x^i$, we require the existence of a symplectic
diffeomorphism $F_{y,x}^*$
such that $d \mu_x e^{2 S_x / \hbar} = F_{y,x}^* ( d \mu_y e^{2 S_y /
\hbar })$. Following Eqn.(4.9) of [\qbi] we know that,
$$F_{y,x}^* (d \mu_y) = {\rho(\psi_y, y)\over\rho(\psi_x, x)} \, \hbox{sdet} \,
\left[ {\partial_l \psi_y \over \partial \psi_x} \right] \cdot d \mu_x\,.
\eqn\sabhalfea$$
so that the background independence condition becomes,
$$\hbox{exp} \, \left({2 S(\psi_x, x)\over\hbar}\right) = \hbox{exp} \, \left(
{2 S(\psi_y, y)\over \hbar}\right) \cdot {\rho(\psi_y, y)\over\rho(\psi_x, x)}
\cdot \, \hbox{sdet} \, \left[ {\p_l \psi_y \over \p \psi_x} \right]\,.
\eqn\sabhalfeb$$
If we consider infinitesimal diffeomorphisms, $y = x + \delta x$, we may
use Eqn.(4.11) of [\qbi],
$$\psi_{x + \delta x}^i = F^i (\psi_x, x, x + \delta x) = \psi^i_x + \delta
x^\mu f^i_\mu (\psi_x , x) + \O (\delta x^2)\,,\eqn\sabhalfec$$
to obtain,
$${\rho(\psi_y, y)\over\rho(\psi_x, x)} \sim 1 + {1\over\rho(\psi_x, x)}
{\p \rho(\psi_x, x) \over\p x^\mu} \, \delta x^\mu + {1\over\rho(\psi_x, x)}
{\p_r \, \rho(\psi_x, x) \over\p \psi^i_x} f^i_\mu \, \delta x^\mu
\,.\eqn\sabhalfec$$
So Eqn.(4.12) of [\qbi] is modified to,
$${\p S(\psi_x, x) \over \p x^\mu} + {\p_r S(\psi_x, x)\over\p \psi^i_x}
f^i_\mu + \half \hbar \left[ {\p \, \hbox{ln} \, \rho\over \p x^\mu} +
{\p_r \, \hbox{ln} \rho \over \p \psi^i_x} f^i_\mu + \, \hbox{str} \, \left(
{\p_l f^i_\mu \over \p \psi^j} \right) \right]\,.\eqn\sabhalfed$$
If we now separate from $f^i_\mu$ the term proportional to the connection
as follows,
$$f^i_\mu \equiv - \wh\Gamma^i_{\mu j} \psi^j - B^i_\mu\,,\eqn\sabhalfee$$
and note that the condition that $F^i$ be a symplectic map reduces to the
condition that there exists an odd Hamiltonian $B_\mu$ such that,
$$B^i_\mu = \omega^{ij} {\p_l B_\mu\over \p \psi^j}\,,\eqn\sabhalfef$$
then it is clear that the new $\psi$-dependence results in a
shift of the $\rho$-dependent term of Eqn.\sabhalfe\ as follows,
$$\eqalign{\p_\mu \, \hbox{ln} \, \rho &\to \p_\mu \, \hbox{ln} \,
\rho - (\, \hbox{ln} \, \rho) \overleftarrow{\p_i} \, \wh\Gamma^i_{\mu j}
\psi^j - (\, \hbox{ln} \, \rho) \overleftarrow{\p_i} \, \omega^{ij} \,
\overrightarrow{\p_j} B_\mu\cr
&\to \p_\mu \, \hbox{ln} \, \rho - (\, \hbox{ln} \, \rho) \,
{\overleftarrow\p\over\p \ket{\Psi}} \, \wh\Gamma_\mu \ket{\Psi} - \{ \,
\hbox{ln} \, \rho , B_\mu \}\cr
&\to D_\mu(\wh\Gamma) \, (\hbox{ln} \, \rho) - \{ \, \hbox{ln} \, \rho ,
B_\mu \}\,.}\eqn\sabhalff$$
where we have made use of Eqns.(4.11), (4.13) and (4.15) of [\qbi]. The
resulting background independence condition is,
$$D_\mu(\wh\Gamma) (S + \half \hbar \, \hbox{ln} \, \rho) = \hbar \wh\Delta
B_\mu + \{ S + \half \hbar \, \hbox{ln} \, \rho , B_\mu \} + \hbar \Delta
\wh\Gamma_\mu\,,\eqn\sabhalfg$$
having used the fact that $\half \hbar \, \hbox{str} \, \wh\Gamma_\mu =
\hbar \Delta
 \wh\Gamma_\mu$, (where $\wh\Gamma_\mu \equiv - \half
\bra{\omega_{12}} \Gamma_\mu^{(2)} \ket{\Psi}_1 \ket{\Psi}_2$).
While the form of the consistency conditions given in
Eqn.\sabhalfg\ is useful for demonstrating the relationship between $\rho$ and
the action, it is, for our purposes, also convenient to write it in the form,
$$\eqalign{D_\mu(\wh\Gamma) S &= \hbar \Delta B_\mu + \{ S , B_\mu \} +
\hbar \Delta \wh\Gamma_\mu - \half \hbar D_\mu(\wh\Gamma) \,
\hbox{ln} \, \rho\cr
&= \Delta_S B_\mu(\wh\Gamma) + \hbar \Delta \wh\Gamma_\mu
- \half \hbar D_\mu(\wh\Gamma) \, \hbox{ln} \, \rho\,.}\eqn\sabhalfl$$
where we have introduced $\Delta_S \, \cdot \equiv \hbar \Delta_{d \mu_S} \,
\cdot \equiv \{ S , \, \cdot \, \} +
\hbar \Delta \, \cdot \, \equiv \{ S + \half \, \hbox{ln} \, \rho , \,
\cdot \, \} + \hbar \wh\Delta \, \cdot \,$. In B-V quantisation
the action is taken to transform as a scalar under string field-redefinitions.
It should therefore be possible to extract any scalar (and in general
field-dependent) component from the action and absorb it into the
B-V density as an additional factor. This procedure should leave the
transformation properties of both the action and $\rho$ unchanged, so that
the requirement that B-V measure $d \mu_S$ be an invariant remains satisfied.
By analysing the transformation properties of Eqn.\sabhalfl\ we will now
verify that it is indeed consistent for $\rho$ to have such field-dependence.

Since the action is a scalar the LHS of Eqn.\sabhalfl, as well
as the first two terms on the RHS (in the first line of the equation) transform
as scalars on the bundle. If the equation is to be consistent, we must
require that the remaining terms also transform correctly. Let us see whether
this is true.

Under a change of basis, $\ket{\Phi_i} \to \ket{\Phi_j} N^j_i(x)$, where
$N^i_j$ is an invertible matrix, we must have $\psi^i \to (N^{-1})^i_j \psi^j$.
We know also that $D_\mu \ket{\Phi_i} \psi^i$ and $\p_\mu$ are also invariants,
and these facts allow us to derive the transformation properties
$\Gamma^j_{\mu i} \to \Gamma'^j_{\mu i}$ of the connection (see [\rsz]),
$$\eqalign{D_\mu \ket{\Phi_i} \psi^i &= \ket{\Phi_i} \p_\mu \psi^i +
\ket{\Phi_j} \Gamma^j_{\mu i} \psi^i\cr
&\to \ket{\Phi} N \p_\mu ( N^{-1} \psi) + \ket{\Phi} N \Gamma'_\mu N^{-1} \psi
\cr
&= \ket{\Phi} \p_\mu \psi + \ket{\Phi} N \p_\mu N^{-1} \psi + \ket{\Phi}
N \Gamma'_\mu N^{-1} \psi\,,}\eqn\sabhalfm$$
where we have adopted matrix notation for brevity. The invariance of
$D_\mu \ket{\Phi_i} \psi^i$ then implies that with our conventions,
$$\Gamma'^{\,i}_{~\mu j} = (N^{-1})^i_{\,k} \, \Gamma^k_{\,\mu l} \, N^l_{\,j}
- (\p_\mu N^{-1})^i_{\,k} \, N^k_{\,j}\,.\eqn\sabhalfn$$
Then the function $\Gamma_\mu = - \half \bra{\omega_{12}} \Gamma^{(2)}_\mu
\ket{\Psi}_1 \ket{\Psi}_2$ associated to the connection transforms as,
$$\Gamma_\mu \to \Gamma_\mu - N (\p_\mu N^{-1})\,.\eqn\sabhalfna$$
The invariance of $d \mu_S$ and the action $S$ implies that the density
must transform according to,
$$\rho \to \rho \, \hbox{sdet} \, \left({\p_l \psi'^j\over\p \psi^i}\right)
= \rho \, \hbox{sdet} \, (N^{-1}) = \rho e^{- \, \hbox{str ln} \, (N^{-1})}
\,.\eqn\sabhalfn$$
Also $\overleftarrow{\p_i} \to \overleftarrow{\p_j} N^j_i$ and therefore,
$$\eqalign{D_\mu \, \hbox{ln} \, \rho &- 2 \Delta \Gamma_\mu = \p_\mu \,
\hbox{ln} \, \rho - \, \hbox{ln} \, \rho \overleftarrow{\p_i} \Gamma^i_{\mu j}
\psi^j - 2 \Delta \Gamma_\mu\cr
&\to \p_\mu \, \hbox{ln} \, \rho - \p_\mu \, \hbox{str ln} \, (N^{-1})
- \, \hbox{ln} \, \rho \overleftarrow{\p} N N^{-1} \Gamma_\mu N N^{-1} \psi
+ \, \hbox{ln} \, \rho \overleftarrow{\p} N (\p_\mu N^{-1}) \psi\cr
&~~~~~
- 2 \Delta \Gamma_\mu + \hbox{str} \, (\p_\mu N^{-1}) N - \hbox{ln} \, \rho
\overleftarrow{\p} N (\p_\mu N^{-1})\psi\cr
&= \p_\mu \, \hbox{ln} \, \rho - \, \hbox{ln} \, \rho \overleftarrow{\p_i}
\Gamma^i_{\mu j} \psi^j - 2 \Delta \Gamma_\mu\,.}\eqn\sabhalfo$$
This verifies that the last two terms on the RHS of Eqn.\sabhalfl\ transform
as scalars, and we conclude that it is indeed consistent to allow $\rho$ some
field-dependence. Nevertheless, given that frames always exist in which $\rho$
may be chosen to be field-independent, we will for simplicity restrict
ourselves to this case in what follows. Choosing $\rho$ to be
field-independent means that we may carry over directly the background
independence condition \sabhalfe, and also set $\Delta = \wh\Delta$.

\section{Origin of General Symplectic Connections}

Until now, we have employed the canonical connection $\wh\Gamma_\mu$. There
is no real reason for restricting ourselves to this particular choice of
connection and for the sake of generality we would like to express the
background independence condition in terms of more general symplectic
connections.

In order to understand how different connections are related to each
other, we will review \S 6.2 of [\qbi] which demonstrates how a
particular choice of connection is related to a choice of three string vertex.

We are interested in the coupling
constant-independent $\O(\hbar)$ terms in the background independence
consistency condition. For the case of field-independent $\rho$, the
$\O(\hbar)$ condition is just Eqn.(6.2) of [\qbi],
$$\p_\mu S_{1,0} = - \half \, \p_\mu \, \hbox{ln} \, \rho + \Delta
\wh\Gamma_\mu + f_\mu(\Delta \wh\B^1_{0,2}) + f_\mu(\V_{1,1})\,.\eqn\sabonebd$$
Now $\wh\B^1_{0,2}$ interpolates from $\I \V_{0,3}$, which is the three string
vertex with one special puncture, to the auxiliary string vertex $\V'_{0,3}$,
so the above equation seems to involve
singular tori $\Delta \V'_{0,3}$. The way this is avoided is to introduce a
new vertex $\wt\V_{0,3}$ with one special puncture and two ordinary
punctures such that $\Delta \wt\V_{0,3}$ is not singular. We can now introduce
a new space $\wt\B^1_{0,2}$ which interpolates between $\V'_{0,3}$ and the
vertex $\wt\V_{0,3}$ and use this to define a new symplectic connection
$\Gamma_\mu(\wt\V_{0,3})$ by,
$$\bra{\omega_{12}} \, \Gamma_\mu^{(1)} = \bra{\omega_{12}} \,
\wh\Gamma_\mu^{(1)} + \int_{\wt\B^1_{0,2}} \bra{\Omega^{(1)0,3}} \wh\O_\mu
\rangle\,,\eqn\sabonebe$$
Absorbing these changes into the canonical connection and $\wh\B^1_{0,2}$, we
make the replacement in Eqn.\sabonebd,
$$\Delta \wh\Gamma_\mu + f_\mu(\Delta \wh\B^1_{0,2}) = \Delta \Gamma_\mu
+ \int_{\Delta({\wh\B^1_{0,2} + \wt\B^1_{0,2}})} \bra{\Omega^{(0)1,1}}
\wh\O_\mu \rangle = \Delta \Gamma_\mu + f_\mu(\Delta \wh\B^1_{0,2} + \Delta
\wt\B^1_{0,2})\,,\eqn\sabonebe$$
The integral is actually path independent, which allows us to define a
moduli space $\B^1_{0,2}(\Gamma)$ satisfying $f_\mu(\Delta \B^1_{0,2}(\Gamma))
= f_\mu(\Delta \wh\B^1_{0,2} + \Delta \wt\B^1_{0,2})$ interpolating from
$\I \V_{0,3}$ to $\wt\V_{0,3}$ in such a way that it completely avoids the
vertex $\V'_{0,3}$, so that the resulting expression avoids any singularities.

We see then that the supertrace of a choice of symplectic connection
$\Gamma_\mu$ is determined by a choice of
three-string vertex $\wt\V_{0,3}$. Alternatively, a choice of symplectic
connection determines $\wt\V_{0,3}$ which in turn allows us to choose
$\B^1_{0,2} (\Gamma)$.

\section{Background Independence and $\Delta_S$-Cohomology}

Having recalled in some detail the origin of generalised connections, we
may now return to our covariant analysis of background independence in terms
of arbitrary connections. As noted in [\moduli], the connection
$\wh\Gamma_\mu$ is a reference symplectic connection and can be shifted as
long as we preserve the symplectic nature of the connection.
Writing the canonical connection in terms of another symplectic connection,
$\wh\Gamma_\mu = \Gamma_\mu - \delta \Gamma_\mu$ we find,
$$D_\mu (\wh\Gamma) S = D_\mu (\wh\Gamma + \delta \Gamma) S +
\{ S , \delta \Gamma_\mu \}\,,\eqn\sabonebb$$
$$\Delta \wh\Gamma_\mu = \Delta (\Gamma_\mu - \delta \Gamma_\mu)
\,,\eqn\sabonebc$$
with $\p_\mu \, \hbox{ln} \, \rho$ invariant. Rearranging terms, we may write
the condition for background independence
Eqn.\sabhalfe\ in terms of a general symplectic connection as,
$$D_\mu (\Gamma) S = \Delta_S B_\mu (\Gamma) +
\hbar \Delta \Gamma_\mu - \half \hbar \p_\mu \, \hbox {ln} \, \rho
\,,\eqn\sabonebcaa$$
where the Hamiltonian $B_\mu(\Gamma)$ for deformations via general
connections is,
$$B_\mu(\Gamma) = B_\mu(\wh\Gamma) - \delta \Gamma_\mu
= B_\mu (\wh\Gamma) - (\Gamma_\mu - \wh\Gamma_\mu)\,.\eqn\sabonebcd$$
Clearly $B_\mu(\Gamma) + \Gamma_\mu$ is invariant under shifts of the
connection, so this final expression actually holds with respect to any
reference connection $\wt\Gamma_\mu$, which may replace the canonical
connection $\wh\Gamma_\mu$.
It follows from \sabonebcaa\ that,
$$\Delta_S (D_\mu(\Gamma) S) = 0\,,\eqn\sabonebcab$$
since both $\Delta \Gamma_\mu$ and $\p_\mu \, \hbox{ln} \, \rho$ are
field-independent. If we now require
uniqueness of the master action, we can use the condition \sabonebcab\ to
derive a cohomology theorem as follows.

Suppose we have a master action $S$ satisfying the master equation. If we now
perturb this action slightly by $\lambda^\mu D_\mu S$, the new action will
also satisfy the master equation,
$$\half \{ S + \lambda^\mu D_\mu S , S + \lambda^\mu D_\mu S \}
+ \hbar \Delta (S + \lambda^\mu D_\mu S) = \lambda^\mu \Delta_S
(D_\mu S) = 0\,.\eqn\saboned$$
We already know from Eqn.\sabonebcab\ that these these marginal deformations
are $\Delta_S$-closed.  The discussion in \S 6 of [\qbi] still applies here,
so that with an appropriate choice of basis we have $\hbox{str}
\, \Gamma_\mu = \p_\mu \, \hbox{ln} \, \rho$, leaving the simplified
condition,
$$D_\mu(\Gamma) S = \Delta_S B_\mu\,,\eqn\sabonedd$$
which tells us that the marginal deformations are also exact. We are left to
conclude that the requirement
of uniqueness of the master action reduces to a cohomology theorem for the
master action in that $D_\mu S$, which being $\Delta_S$-closed, must also be
$\Delta_S$-exact. We will come back to this point in more detail in \S 5.
Let us now examine the commutator of deformations.

\chapter{\bf The Commutator of Background Deformations}

In this section we will take a second covariant derivative of
\sabonebcaa\ and demonstrate the existence of a $\Delta_S$-closed `field
strength' $H_{\mu\nu}$, which in its turn will imply the existence of a
Hamiltonian $B_{\mu\nu}$ from uniqueness of the master action.

\section{The Commutator Conditions}

\noindent
We shall begin with Eqn.\sabonebcaa\ which expresses the background
independence condition in terms of a general symplectic connection,
$$D_\mu (\Gamma) S = \Delta_S B_\mu (\Gamma) + \hbar \Delta \Gamma_\mu
- \half \hbar \p_\mu \, \hbox{ln} \, \rho\,.\eqn\sabonej$$
Before proceeding, we derive the useful identity $[ D_\mu ,\Delta ] F = 0$
(for arbitrary functions $F$) which holds if the connection is symplectic.
From their definitions we have,
$$D_\mu F \equiv \p_\mu F - F \overleftarrow{\p_i} \Gamma^i_{\mu j} \psi^j
= \p_\mu F - \Gamma^i_{\mu j} \psi^j \overrightarrow{\p_i} F\,,\eqn\saboneja$$
$$\Delta F \equiv \half (-)^i \overrightarrow{\p_i} ( \omega^{ij}
\overrightarrow{\p_j} F ) = \half (-)^{i+ij} \omega^{ij}
\overrightarrow{\p_i} \overrightarrow{\p_j} F\,.\eqn\sabonejb$$
The commutator is calculated thus,
$$\eqalign{[ D_\mu , \Delta ] F &= \p_\mu ( \half (-)^{i+ij} \omega^{ij}
\overrightarrow{\p_i} \overrightarrow{\p_j} F) - \Gamma^k_{\mu l} \psi^l
\overrightarrow{\p_k} ( \half (-)^{i+ij} \omega^{ij} \overrightarrow{\p_i}
\overrightarrow{\p_j} F )\cr
&~~~~~~~~~~ - \half (-)^{i+ij} \omega^{ij}
\overrightarrow{\p_i} \overrightarrow{\p_j} ( \p_\mu F - \Gamma^k_{\mu l}
\psi^l \overrightarrow{\p_k} F )\cr
&= \half(-)^{i+ij} \bigl( \p_\mu \omega^{ij} \overrightarrow{\p_i}
\overrightarrow{\p_j} F - (-)^k \Gamma^k_{\mu l}
\psi^l \omega^{ij} \overrightarrow{\p_i} \overrightarrow{\p_j}
\overrightarrow{\p_k} F\cr
&~~~~~~~~~~ + (-)^{k+l} \Gamma^k_{\mu l} \omega^{ij}
\overrightarrow{\p_i} \overrightarrow{\p_j} ( \psi^l \overrightarrow{\p_k} F )
\bigr)\cr
&= \half(-)^{i+ij} \bigl( \p_\mu \omega^{ij} \overrightarrow{\p_i}
\overrightarrow{\p_j} F + (-)^{j+k} \Gamma^k_{\mu j} \omega^{ij}
\overrightarrow{\p_i} \overrightarrow{\p_k} F + (-)^{ij+i+k} \Gamma^k_{\mu i}
\omega^{ij} \overrightarrow{\p_j} \overrightarrow{\p_k} F \bigr)\cr
&= \half(-)^{i+ij} \bigl( \p_\mu \omega^{ij} + \Gamma^i_{\mu k}
\omega^{kj} - (-)^{(i+1)(j+1)} \Gamma^j_{\mu k} \omega^{ki}  \bigr)
\overrightarrow{\p_i} \overrightarrow{\p_j} F \equiv 0\,,}\eqn\sabonejc$$
as the vanishing of the expression in brackets is precisely the condition for
the connection to be symplectic. (Note that the last two terms in the brackets
are actually equal, though we have chosen to separate the terms as above in
order to make the symplectic identity explicit).

\noindent
Another result we will need is the following,
$$\eqalign{\Delta R_{\mu\nu} &= \half (-)^i ( \p_\mu \Gamma^i_{\nu i}
- \p_\nu \Gamma^i_{\mu i} + \Gamma^i_{\mu j} \Gamma^j_{\nu i} -
\Gamma^i_{\nu j} \Gamma^j_{\mu i} )\cr
&= \half (-)^i ( \p_\mu \Gamma^i_{\nu i} - \p_\nu \Gamma^i_{\mu i} )\cr
&= \p_\mu \Delta \Gamma_\nu - \p_\nu \Delta \Gamma_\mu\cr
&= D_\mu \Delta \Gamma_\nu - D_\nu \Delta \Gamma_\mu\,,}\eqn\sabonek$$
where we have have used the fact that $\Delta \Gamma_\mu$ is
field-independent.

\noindent
Additionally, the field-independence of $\rho$ allows us to ignore its mixed
covariant derivatives,
$$[ D_\mu , D_\nu ] \, \hbox{ln} \, \rho = [ \p_\mu , \p_\nu ] \, \hbox{ln} \,
\rho = 0\,.\eqn\sabonel$$
We are now ready to take a second covariant derivative of Eqn.\sabonej.
Making the connection $\Gamma_\mu$ implicit we have,
$$\eqalign{[ D_\mu , D_\nu ] S &= D_\mu D_\nu S - D_\nu D_\mu S\cr
&= D_\mu \Delta_S B_\nu - D_\nu \Delta_S B_\mu + \hbar ( D_\mu \Delta
\Gamma_\nu - D_\nu \Delta \Gamma_\mu )\cr
&= D_\mu \{ S , B_\nu \} - D_\nu \{ S , B_\mu \} + \hbar ( D_\mu \Delta B_\nu -
D_\nu \Delta B_\mu) + \hbar \Delta R_{\mu\nu}\cr
&= \{ D_\mu S , B_\nu \} - \{ D_\nu S , B_\mu \} + \{ S , D_\mu B_\nu - D_\nu
B_\mu \}\cr
&\,\,\,\,\,\,\,\,\,\,
+ \hbar \Delta (D_\mu B_\nu - D_\nu B_\mu) + \hbar( [ D_\mu , \Delta ] B_\nu -
[ D_\nu , \Delta ] B_\mu) + \hbar \Delta R_{\mu\nu}\,.}\eqn\sabonem$$
where we have made use of Eqns.\sabonek\ and \sabonel. The result proven above
allows us to discard the commutators $[ D_\mu , \Delta ]$ so that,
$$\eqalign{[ D_\mu , D_\nu ] S &= \{ \{ S , B_\mu \} + \hbar \Delta B_\mu ,
B_\nu \} - \{ \{ S , B_\nu \} + \hbar \Delta B_\nu , B_\mu \}\cr
&\,\,\,\,\,\,\,\,\,\,\,\,\,\,\,\,\,\,\,\,
+ \Delta_S (D_\mu B_\nu - D_\nu B_\mu) + \hbar \Delta R_{\mu\nu}\cr
&= \{ \{ S , B_\mu \} , B_\nu \} - \{ \{ S , B_\nu \} , B_\mu \} + \hbar
\{ \Delta B_\mu , B_\nu\} - \hbar \{ \Delta B_\nu , B_\mu \}\cr
&\,\,\,\,\,\,\,\,\,\,\,\,\,\,\,\,\,\,\,\, + \Delta_S
(D_\mu B_\nu - D_\nu B_\mu) + \hbar \Delta R_{\mu\nu}\cr
&= \{ S , \{ B_\mu , B_\nu \} \} + \hbar \Delta \{ B_\mu , B_\nu \}
+ \Delta_S (D_\mu B_\nu - D_\nu B_\mu) + \hbar \Delta R_{\mu\nu}\cr
&= \Delta_S (\{ B_\mu , B_\nu \} + D_\mu B_\nu - D_\nu B_\mu)
+ \hbar \Delta R_{\mu\nu}\,.}\eqn\sabseven$$
But we know that the action of the commutator is related to the antibracket
with the curvature,
$$\eqalign{[ D_\mu , D_\nu ] S &= - \{ S , R_{\mu\nu} \}\cr
&= - \Delta_S R_{\mu\nu} + \hbar \Delta R_{\mu\nu}\,.}\eqn\sabeight$$
Combining Eqns.\sabseven\ and \sabeight\ we find the consistency condition,
$$\Delta_S H_{\mu\nu} = 0\,,\eqn\sabnine$$
where we have introduced the field strength,
$$H_{\mu\nu} \equiv \{ B_\mu , B_\nu \} + D_\mu B_\nu - D_\nu B_\mu
+ R_{\mu\nu}\,.\eqn\sabninea$$
Of course this equation must be satisfied since the condition results
directly from Eqn.\sabonebcaa, which has already been solved explicitly.
Eqn.\sabnine\ implies that a perturbed master action
$S' = S + \lambda^{\mu\nu} H_{\mu\nu}$ also satisfies the master equation,
$$\eqalign{\half \{ S' , S' \} + \hbar \Delta S' &= 
\half \{ S + \lambda^{\mu\nu} H_{\mu\nu}, S + \lambda^{\mu\nu}
H_{\mu\nu} \} + \hbar \Delta (S + \lambda^{\mu\nu} H_{\mu\nu})\cr
&= \half \{ S , S \} + \hbar \Delta S + \{ S , \lambda^{\mu\nu}
H_{\mu\nu} \} + \hbar \Delta \lambda^{\mu\nu} H_{\mu\nu}\cr
&= \lambda^{\mu\nu} \{ S , H_{\mu\nu} \} + \hbar \lambda^{\mu\nu} \Delta
H_{\mu\nu}\cr
&= \lambda^{\mu\nu} \Delta_S H_{\mu\nu}\cr
&= 0\,.}\eqn\sabten$$
The hope that the master action be unique up to gauge transformations would
require that the perturbed action be merely a field redefined version of the
original. This is so if there exists a Hamiltonian $B_{\mu\nu}$ such that,
$$H_{\mu\nu} = \Delta_S B_{\mu\nu}\,.\eqn\sabeleven$$
So we see that the existence of $B_{\mu\nu}$ or alternatively, uniqueness of
the string action, implies a (higher) cohomology theorem for the string
action which in turn implies quantum background independence of the string
action with respect to commutators of deformations.

\section{Analysis of Gauge Freedom of $B_{\mu\nu}$}

\noindent
Having postulated the existence of the object $B_{\mu\nu}$, let us now explore
the extent to which it is uniquely defined.

It was shown for the classical case in [\moduli] that shifting the connection
whilst retaining the symplectic property does not alter $H_{\mu\nu}$.
An identical argument which we need not repeat here shows that this
statement also holds in the quantum case, so that $B_{\mu\nu}$ does not
depend on the particular choice of symplectic connection.

From the nilpotency of $\Delta_S$, any shift of $B_\mu$ by a $\Delta_S$-trivial
object will clearly also satisfy the background independence condition
Eqn.\sabonebcaa,
$$B_\mu \, \to \, B_\mu + \Delta_S \lambda_\mu\,.\eqn\sabtwelvea$$
This results in a corresponding shift in $H_{\mu\nu}$ given by,
$$\eqalign{H_{\mu\nu} &\to H_{\mu\nu} + \{ \Delta_S \lambda_\mu , B_\nu \}
+ \{ B_\mu , \Delta_S \lambda_\nu \} + D_\mu (\Delta_S \lambda_\nu) -
D_\nu (\Delta_S \lambda_\mu)\cr
&\to H_{\mu\nu} + \Delta_S \{ \lambda_\mu , B_\nu \} + \{ \lambda_\mu ,
\Delta_S B_\nu \} + \Delta_S \{ B_\mu , \lambda_\nu \} - \{ \Delta_S B_\mu ,
\lambda_\nu \}\cr
&\,\,\,\,\,\,\,\,\,\,\,\,\,\,\,\,\,\,\,\,
+ D_\mu (\{ S , \lambda_\nu \} + \hbar \Delta \lambda_\nu) -
D_\nu (\{ S , \lambda_\mu \} + \hbar \Delta \lambda_\mu)\cr
&\to H_{\mu\nu} + \Delta_S (\{ B_\mu , \lambda_\nu \} - \{ B_\nu ,
\lambda_\mu \}) - \{ \Delta_S B_\mu , \lambda_\nu \} + \{ \Delta_S B_\nu,
\lambda_\mu \}\cr
&\,\,\,\,\,\,\,\,\,\,\,\,\,\,\,\,\,\,\,\,
+\hbar(D_\mu\Delta\lambda_\nu - D_\nu\Delta\lambda_\mu)
+\{D_\mu S,\lambda_\nu\} + \{ S,D_\mu\lambda_\nu\} - \{D_\nu S,
\lambda_\mu\} - \{S,D_\nu\lambda_\mu\}\,.}\eqn\sabthirteena$$
We can now use Eqn.\sabonej\ and the fact that $D_\mu$ and $\Delta$ commute,
$$\eqalign{H_{\mu\nu} &\to
H_{\mu\nu} + \Delta_S(\{B_\mu,\lambda_\nu\} - \{B_\nu,\lambda_\mu\})
+ \{S,D_\mu\lambda_\nu-D_\nu\lambda_\mu\} + \hbar\Delta(D_\mu\lambda_\nu
- D_\nu\lambda_\mu)\cr
&\,\,\,\,\,\,\,\,\,\,\,\,\,\,\,\,\,\,\,\,
+ \{ \Delta_S B_\mu , \lambda_\nu \} - \{ \Delta_S B_\nu ,
\lambda_\mu \} - \{ \Delta_S B_\mu , \lambda_\nu \} + \{ \Delta_S B_\nu ,
\lambda_\mu \}\cr
&\to H_{\mu\nu} + \Delta_S (\{ B_\mu , \lambda_\nu \} - \{ B_\nu ,
\lambda_\mu\} + D_\mu \lambda_\nu - D_\nu \lambda_\mu)\cr
&\to H_{\mu\nu} + \Delta_S ({\cal D}_\mu \lambda_\nu - {\cal D}_\nu
\lambda_\mu)\,,}\eqn\sabthirteen$$
where ${\cal D}_\mu \equiv \{ B_\mu , \, \cdot \, \} + D_\mu$ is the
`gauge covariant derivative' introduced in [\moduli]. Given that we seek
$B_{\mu\nu}$ such that $H_{\mu\nu} = \Delta_S B_{\mu\nu}$, the shift $B_\mu
\to B_\mu + \Delta_S \lambda_\mu$ must
correspond to a non-trivial gauge freedom,
$$B_{\mu\nu} \,\to \, B_{\mu\nu} + \D_\mu \lambda_\nu - 
\D_\nu \lambda_\mu\,.\eqn\sabfourteen$$
There is also of course a trivial gauge freedom under $B_{\mu\nu} \to
B_{\mu\nu} + \Delta_S \lambda_{\mu\nu}$. We will give a detailed
interpretation of these in the sequel.

\section{Consistency Conditions and Recursion Relations for Moduli Spaces}

\noindent
We will now examine explicitly the consistency conditions
derived in \S 4.1,
$$D_\mu B_\nu - D_\nu B_\mu + \{ B_\mu , B_\nu \} + R_{\mu\nu} =
\Delta_S B_{\mu\nu}\,.\eqn\sabfifteen$$
The aim of the present section is to show that the Hamiltonian $B_{\mu\nu}$
is the function associated to some moduli spaces of Riemann surfaces of
any genus and with two special punctures. We will follow the analysis of
[\moduli] to derive the recursion relations which must be satisfied by the
higher genus moduli spaces that define the Hamiltonian $B_{\mu\nu}$.

\noindent
We therefore take $B_{\mu\nu}$ to be a Hamiltonian of the form,
$$B_{\mu\nu} = - f_{\mu\nu} (\B^2) = - \int_{\B^2} \bra{\Omega_{\bar 1\bar 2 }}
\, \wh\O_\mu\rangle_{\bar 1} \ket{ \wh\O_\nu}_{\bar 2}\,,\eqn\sabsixteen$$
where $\B^2$ is a sum of moduli spaces of surfaces with two special
punctures, now extended to include higher genus terms.

\noindent
In terms of moduli spaces the right hand side of Eqn.\sabfifteen\ may be
written,
$$\eqalign{\Delta_S B_{\mu\nu} &= - \{ S , f_{\mu\nu} (\B^2) \} -
\hbar \Delta f_{\mu\nu} (\B^2) \cr
&= - \{ Q + f(\V) + \hbar S_{1,0} , f_{\mu\nu} (\B^2) \} +
\hbar f_{\mu\nu} (\Delta \B^2)\cr
&= f_{\mu\nu} (\p \B^2 + \{ \V , \B \} + \hbar \Delta \B^2)\cr
&= f_{\mu\nu} (\delta_\V \B^2)\,.} \eqn\sabseventeen$$
Recalling (from \S 3.2 of [\moduli]) that the left hand side of \sabfifteen\ is
independent of the
connection, we may simply apply the results derived in \S 6 of [\moduli]
for the canonical connection (noting of course that $\B^1$ now includes the
spaces of higher genus) which tells us that,
$$D_\mu B_\nu  - D_\nu B_\mu + \{B_\mu , B_\nu\} + R_{\mu\nu} = f_{\mu\nu}
(\T^2_{0,1} + (\K - \I) \B^1 - \half \{ \B^1 , \B^1 \} )\,. \eqn\sabeighteen$$
Putting these together, Eqn.\sabfifteen\ becomes,
$$f_{\mu\nu} (\T^2_{0,1} + (\K - \I) \B^1 - \half \{ \B^1 , \B^1 \}) =
f_{\mu\nu} (\p_\V \B^2)\,.\eqn\sabnineteen$$
This equation will be satisfied if,
$$\delta_\V \B^2 = \T^2_{0,1} + (\K - \I) \B^1 - \half \{ \B^1 , \B^1 \}\,.
\eqn\sabtwenty$$
This has the same form as the classical formula, except that
$\delta_\V$ now contains in its definition the additional operator
$\hbar \Delta$.

Let us verify the consistency of Eqn.\sabtwenty\ by checking that $\delta_\V$
acting on the RHS vanishes. Consider each term separately we find,
$$\delta_\V \T^2_{0,1} = \p \T^2_{0,1} + \{ \V , \T^2_{0,1} \} +
\hbar \Delta \T^2_{0,1} = \I \V'_{0,3} + \{ \V , \T^2_{0,1} \}\,,
\eqn\sabtwentya$$
$$\eqalign{\delta_\V& (\K - \I) \B^1 = [ \delta_\V , \K - \I ] \B^1 +
(\K - \I) \delta_\V \B^1\cr
&= \{ \V'_{0,3} + (\K - \I) \V , \B^1 \} + (\K - \I) (\V'_{0,3} + (\K - \I) \V
+ \hbar \Delta \B^1_{0,2} + \hbar \I \V_{1,1})\cr
&= \{ \V'_{0,3} + (\K - \I) \V , \B^1 \} - \I \V'_{0,3} - \{ \V , \T^2_{0,1} \}
+ \hbar \K \Delta \B^1_{0,2} + \hbar \K \I \V_{1,1}\,,}\eqn\sabtwentyone$$
$$\delta_\V (- \half \{ \B^1 , \B^1 \}) = - \{ \delta_\V \B^1 , \B^1 \} = 
- \{ \V'_{0,3} + (\K - \I) \V , \B^1 \}\eqn\sabtwentyonea$$
We remind ourselves from the discussion of \S 3.3 that $\B^1_{0,2}$
interpolates between $\I\V_{0,3}$ and some vertex $\wt\V_{0,3}$ determined by
the choice of connection,
so that there is no longer the unwanted singularity
associated with $\Delta \V'_{0,3}$. So Eqn.\sabtwenty\ is consistent if,
$$\K (\Delta \B^1_{0,2} + \I \V_{1,1}) = 0\,.\eqn\sabtwentytwo$$
Both terms in this expression consists of the operator $\K$ acting on a
torus or tori with a single special puncture. It is fairly simple to see
why each term must vanish. 

Consider a torus $T^1_{1,0}$ with a single special puncture. The operator $\K$
adds  another special puncture over the remainder of the surface of the torus,
antisymmetrising with respect to the two punctures. The translational
symmetry of the torus means that for any relative position of the two
punctures of any torus in $\K T^1_{1,0}$ there will be another torus with
the two punctures with positions reversed. The antisymmetrising property of
$\K$ ensures that this pair of twice-punctured tori will occur with the
opposite sign and cancel. This pairwise cancellation means that each term
in Eqn.\sabtwentytwo\ vanishes, thus verifying the consistency of
Eqn.\sabtwenty and thereby Eqn.\sabsixteen.

\chapter{\bf $\Delta_S$-Cohomology Classes and Theory Space Geometry}

Having introduced $\B$-spaces with two special punctures in the previous
section, we will outline in this section how uniqueness of the master action
implies the existence of $\B$-spaces with more than two punctures. We will
show that the quantum generalisation of the analysis [\nonconf] has an
efficient description in terms of differential forms on the theory
space manifold which is parametrised by the marginal operators.

Let us consider a basis of marginal states, $\{ \ket{\wh\O_1}, . \, . \, . \, ,
\ket{\wh\O_N} \}$,
where $N$ may be infinite. For a given point $S$ in the theory space manifold,
these form a basis of the tangent space T$_*M_S$. This $N$-dimensional theory
space manifold $M_S$ may therefore be parametrised locally by coordinates
$\{ x^1, . \, . \, . \, , x^N\}$.
We will use the following notation for differential forms on T$^*M_S$,
$${\bf A}_{(n)} \equiv {1\over n!} A_{\mu_1\cdots\mu_n} {\bf d}x^{\mu_1}
\wedge \cdots \wedge {\bf d}x^{\mu_n}\,.\eqn\sabthirtyfour$$
In this language the $B_\mu$ are
components of a one-form `gauge field' ${\bf B}_{(1)} = B_\mu {\bf d}x^\mu$,
and the $H_{\mu\nu}$ are components of a two-form field strength
${\bf H}_{(2)}  = \half H_{\mu\nu} {\bf d}x^\mu \wedge {\bf d}x^\nu$ such that
${\bf H}_{(2)}  = \half \Delta_S
{\bf B}_{(2)} = \Delta_S B_{\mu\nu} {\bf d}x^\mu \wedge {\bf d}x^\nu$.

\noindent
It is convenient also to introduce a kind of
`gauge-covariant exterior derivative' ${\bf\D = D + \{ B , \, \cdot \, \}}$
(not to be confused with the actual exterior derivative ${\bf d}$ on
T$^*M_S$), which is defined by,
$$\eqalign{{\bf\D A}_{(n)} &\equiv {\bf D A}_{(n)}+\{ {\bf B}_{(1)} ,
{\bf A}_{(n)} \}\cr
&= {1\over n!} \D_{\mu_0} A_{\mu_1\cdots\mu_n} {\bf d}x^{\mu_0} \wedge \cdots
\wedge {\bf d}x^{\mu_n}\cr
&= {1\over n!} (D_{\mu_{0}} A_{\mu_1\cdots\mu_n} + \{ B_{\mu_{0}} ,
A_{\mu_1\cdots\mu_n} \}) {\bf d}x^{\mu_{0}} \wedge \cdots \wedge
{\bf d}x^{\mu_n}\,.}\eqn\sabthirtyfouraa$$
If we define the antibracket of two forms by,
$$\{ {\bf A}_{(n)} , {\bf C}_{(m)} \} = {1\over n! m!} \{ A_{\mu_1\cdots\mu_n}
, C_{\mu_{n+1}
\cdots\mu_{n+m}} \} {\bf d}x^{\mu_1} \wedge \cdots \wedge {\bf d}x^{\mu_n}
\wedge {\bf d}x^{\mu_n+1} \wedge \cdots \wedge {\bf d}x^{\mu_{n+m}}\,,\eqn
\sabthirtyfoura$$
then ${\bf\D}$ has the property,
$${\bf\D} \{ {\bf A}_{(n)} , {\bf C}_{(m)} \} = \{ {\bf\D} {\bf A}_{(n)} ,
{\bf C}_{(m)} \} + (-)^n \{ {\bf A}_{(n)} , {\bf\D} {\bf C}_{(m)} \}\,.\eqn
\sabthirtyfourb$$
As a useful identity, we show that $\D$ commutes with $\Delta_S$,
$$\eqalign{[ \Delta_S , {\bf\D ] A}_{(n)} &= \Delta ({\bf D A}_{(n)} +
\{ {\bf B}_{(1)} , {\bf A}_{(n)} \}) + \{ S , {\bf D A}_{(n)} + \{
{\bf B}_{(1)} , {\bf A}_{(n)}
\} \} - {\bf\D} \Delta_S {\bf A}_{(n)}\cr
&= {\bf D} \Delta {\bf A}_{(n)} + \{ \Delta {\bf B}_{(1)} , {\bf A}_{(n)} \} +
\{ {\bf B}_{(1)} , \Delta {\bf A}_{(n)} \} + {\bf D} \{ S , {\bf A}_{(n)}
\}\cr
&\,\,\,\,\,\,\,\,\,\,- \{ {\bf D} S , {\bf A}_{(n)} \} + \{ {\bf B}_{(1)}
, \{ S , {\bf A}_{(n)} \} \} + \{ \{ S , {\bf B}_{(1)} \} ,
{\bf A}_{(n)} \} - {\bf\D} \Delta_S {\bf A}_{(n)}\cr
&= {\bf\D} \Delta_S {\bf A}_{(n)} - \{ {\bf D} S , {\bf A}_{(n)} \} +
\{ \Delta_S {\bf B}_{(1)} , {\bf A }_{(n)} \} - {\bf\D} \Delta_S
{\bf A}_{(n)}\cr
&= 0\,.}\eqn\sabthirtyfourc$$
Another useful property is the following,
$$\eqalign{{\bf\D \D A}_{(n)} &= {1\over n!} \D_{\mu_a} \D_{\mu_b}
A_{\mu_1\cdots\mu_n}
{\bf d}x^{\mu_a} \wedge {\bf d}x^{\mu_b} \wedge {\bf d}x^{\mu_1} \wedge \cdots
\wedge {\bf d}x^{\mu_n}\cr
&= {1\over 2n!} [ \D_{\mu_a} , \D_{\mu_b} ] A_{\mu_1\cdots\mu_n}
{\bf d}x^{\mu_a}
\wedge {\bf d}x^{\mu_b} \wedge {\bf d}x^{\mu_1} \wedge \cdots \wedge
{\bf d}x^{\mu_n}\cr
&= {1\over 2n!} \{ H_{\mu_a\mu_b} , A_{\mu_1\cdots\mu_n} \} {\bf d}x^{\mu_a}
\wedge {\bf d}x^{\mu_b} \wedge {\bf d}x^{\mu_1} \wedge \cdots \wedge
{\bf d}x^{\mu_n}\cr
&= \{ {\bf H}_{(2)} , {\bf A}_{(n)} \}\,,}\eqn\sabthirtyfourd$$
where we have made use of the identity $[ \D_\mu , \D_\nu ] \, \cdot \, =
\{ H_{\mu\nu} , \, \cdot \, \}$ (note the sign correction to Refs.[\moduli]
and [\nonconf]).

Let us now proceed to show how to recursively construct in a simple
manner the $n$-form field strengths ${\bf H}_{(n)}$ and gauge fields
${\bf B}_{(n)}$ for all $n \geq 2$. Treating the modified action ${\bf \bar S}
\equiv {\bf S } + \half \hbar \, \hbox{ln} \, {\bf \rho}$
as a zero-form and $B_\mu$ and $\Gamma_\mu$ as components of one-forms, we
may write the background independence conditions Eqn.\sabonebcaa\ as,
$${\bf\D \bar S} = \Delta ({\bf B}_{(1)} + {\bf \Gamma}_{(1)})\,.\eqn
\sabthirtyfoure$$
Acting once again with the gauge-covariant exterior derivative,
$$\eqalign{{\bf \D \D \bar S} &= \D_\mu \Delta (B_\nu + \Gamma_\nu)
{\bf d}x^\mu \wedge {\bf d}x^\nu\cr
&= \Delta {\bf H}_{(2)}\,.}\eqn\sabthirtyfourf$$
But we know from Eqn.\sabthirtyfourd\ that ${\bf\D\D \bar S} = \{ {\bf H}_{(2)}
, {\bf\bar S} \}$, from which immediately follows the result we derived
earlier (now written in terms of forms),
$$\Delta_S {\bf H}_{(2)} = 0\,.\eqn\sabthirtyfourg$$
Now, by the same argument which was used in Eqn.\saboned, we know that we can
add to the action any $\Delta_S$-closed function to get a new action also
satisfying the master equation. The hope that the master action be unique
implies no non-trivial $\Delta_S$-cohomology, which leads us naturally to the
requirement that ${\bf H}_{(2)}$ be $\Delta_S$-exact,
$${\bf H}_{(2)} = \Delta_S {\bf B}_{(2)}\,.\eqn\sabthirtythree$$
We have already shown that the Hamiltonian $B_{\mu\nu}$ may be obtained from
moduli spaces of surfaces with two special punctures.

The procedure to construct higher forms goes as follows. Let us define an
auxiliary three-form,
$${\bf H}'_{(3)} = {\bf\D B}_{(2)}\,.\eqn\sabthirtythreea$$
Acting on this with $\Delta_S$, we find,
$$\Delta_S {\bf H}'_{(3)} = \Delta_S {\bf\D B}_{(2)} = {\bf\D} \Delta_S
{\bf B}_{(2)} = {\bf\D H}_{(2)} = 0\,,\eqn\sabthirtythreeb$$
where we have used Eqns.\sabthirtyfourc, \sabthirtythree\ and finally the
Bianchi identity for $H_{\mu\nu}$, Eqn.(2.26) of [\nonconf].
This means we can simply choose our $\Delta_S$-closed three-form field
strength to be,
$${\bf H}_{(3)} \equiv {\bf H}'_{(3)} = {\bf\D B}_{(2)}\,,\eqn
\sabthirtythreec$$
which is a condensed way of expressing the analogous classical result
Eqn.(2.27) of
[\nonconf]. Once again, uniqueness of the master action requires the existence
of a corresponding three-form gauge field such that,
$${\bf H}_{(3)} = \Delta_S {\bf B}_{(3)}\,.\eqn\sabthirtythreed$$
Finding the four-form field strength is still simple, albeit not quite as
trivial. We first define an auxiliary four-form,
$${\bf H}'_{(4)} = {\bf\D B}_{(3)}\,.\eqn\sabthirtythreee$$
Acting with $\Delta_S$ gives a long chain of identities,
$$\eqalign{\Delta_S {\bf H}'_{(4)} &= \Delta_S {\bf\D B}_{(3)} = {\bf\D}
\Delta_S {\bf B}_{(3)} = {\bf\D H}_{(3)} = {\bf\D \D B}_{(2)}\cr
&= \{ {\bf H}_{(2)} , {\bf B}_{(2)} \} = \{ \Delta_S {\bf B}_{(2)} ,
{\bf B}_{(2)} \} = \half \Delta_S \{ {\bf B}_{(2)} , {\bf B}_{(2)} \}
\,.}\eqn\sabthirtythreeb$$
From this we can extract the $\Delta_S$-closed four-form field strength,
$${\bf H}_{(4)} \equiv \D {\bf B}_{(3)} - \half \{ {\bf B}_{(2)} ,
{\bf B}_{(2)} \}\,.\eqn\sabthirtythreec$$
Note that this simplified expression agrees with Eqn.(2.33) of [\nonconf].
By repeating the same procedure that is, defining an auxiliary $p$-form
by ${\bf H}'_{(p)} = \D {\bf B}_{(p-1)}$, and then acting upon it with
$\Delta_S$ to eventually extract a $\Delta_S$-closed $p$-form ${\bf H}_{(p)}$,
we may construct all higher $n$-forms ${\bf H}_{(n)} = \Delta_S {\bf B}_{(n)}$
{\it ad nauseam}. We will refer to this last equality as the $n$-th vanishing
theorem for $\Delta_S$ cohomology classes. Indeed one can shown by
induction that the general formula is,
$${\bf H}_{(n)} = {\bf\D B}_{(n-1)} + \half \sum_{m=2}^{n-2} (-)^{m+1} \{
{\bf B}_{(m)} , {\bf B}_{(n-m)} \}~~~~~~~~~~(n>2)\,.\eqn\sabthirtythreed$$

Generalising the results which have already been demonstrated for $B_\mu$ and
$B_{\mu\nu}$, we can assume that the antisymmetric coefficients of the
$n$-form Hamiltonians ${\bf B}_{(n)}$ are given by functions (with the
appropriate marginal state insertions), of moduli spaces with $n$ special
punctures,
$$B_{\mu_1\cdots\mu_n} = - f_{\mu_1\cdots\mu_n} (\B^n) = - {1\over n! \bar n!}
\int_{\B^n} \bra{\Omega_{\bar 1\cdots\bar n}}\wh\O_{\mu_1}\rangle_{\bar 1}
\cdots\ket{\wh\O_{\mu_n}}_{\bar n}\,.\eqn\sabthirtysix$$
We note that $\B^n = \sum_{g,k\geq0} \B^n_{g,k}$ extends over a complete set
of positive-dimensional moduli spaces of punctured Riemann surfaces for all
genera, and all numbers of ordinary punctures compatible with the
dimensionality requirement.

Just as before, these $\B^n$-spaces may be explicitly constructed using their
recursion relations, but we should defer this task until the sequel [\geom],
when we have before us the complete $\B$-complex and corresponding recursion
relations.

In summary, what he have shown is that the requirement of unique physics
implies the need for background independence. The statement of uniqueness
and background independence at the $p$-th order of deformations implies the
$p$-th vanishing theorem for the $\Delta_S$ cohomology class of the master
action, so that uniqueness and background independence to all orders implies
a set of $\Delta_S$ cohomology vanishing theorems for the master action $S$.
In addition, the
space of equivalent theories related by marginal deformations is simply the
equivalence class (contained in the manifold $M_S$) of the
$\Delta_S$-cohomology of which the action $S$ is a representative.

\chapter{\bf Conclusion}

In this paper we have generalised the work [\moduli] of Zwiebach from the
classical to the quantum case, showing that uniqueness and background
independence of the master action to all orders implies a set of
cohomology vanishing theorems for the closed string action, and have
postulated the
existence of all punctured higher genus interpolating moduli spaces
$\B^{\bar n}_{g,n}$ of positive dimension.

\ack
I am indebted to my supervisor Barton Zwiebach without whom the present work
would not have existed. I would also like to thank Mohiuddin Khan Khokon for
his support and  encouragement, and Chris Isham for his hospitality during
my stay at Imperial College.

\appendix

We collect here for reference some useful formulae used in this paper. We
assume that the states $\ket{\O_\mu}$ are BRST-closed.

$$\{ \A , \B \} = -(-)^{(\A+\bar n_\A+1) (\B+\bar n_\B+1)} \{ \B , \A \}\,,
\eqn\sabbyseventeen$$
$$(-)^{(\A+\bar n_\A+1)(\C+\bar n_\C+1)} \{ \{ \A , \B \} , \C \} + cycl.
= 0\,,\eqn\sabbyeighteen$$
$$(-)^{(\A+\bar n_\A+1)(\C+\bar n_\C+1)} \{ \A , \{ \B , \C \} \} + cycl.
= 0\,,\eqn\sabbynineteen$$
$$\Delta \{ \A , \B \} = \{ \Delta \A , \B \} + (-)^{\A+\bar n_\A+1} \{ \A ,
\Delta \B \}\,,\eqn\sabbytwentyone$$
$$\Delta_S^2 = 0\,,\eqn\sabbyone$$
$$\Delta_S \{ \A , \B \} = \{ \Delta_S \A , \B \} + (-)^{\A+\bar n_\A+1} \{
\A , \Delta_S \B \}\,,\eqn\sabbytwentyone$$
$$D_\mu \{ \A , \B \} = \{ D_\mu \A , \B \} + \{ \A , D_\mu \B \}
\,,\eqn\sabbytwentyone$$
$$f_{\mu\nu} (\{ \V , \A \}) = - \{ f(\V) , f_{\mu\nu}(\A) \}\,,\eqn\sabbytwo$$
$$\delta_\V \delta_\V \A = 0\,,\eqn\sabbythirtythree$$
$$[ \delta_\V , \K ] \A = (-)^{\A+\bar n_\A} \{ \V'_{0,3} + \K \V, \A \}
\,,\eqn\sabbythirtyfour$$
$$[ \delta_\V , \I ] \A = (-)^{\A+\bar n_\A} \{ \I \V , \A \}\,.
\eqn\sabbythirtyfive$$
$$\K \K \A = 0\,,\eqn\sabbythirtysix$$
$$(\K \I + \I \K) \A = \{ \A , \T^2_{0,1} \}\,.\eqn\sabbythirtyseven$$
$$\I \I \A = 0\,.\eqn\sabbythirtyeight$$
$$\p \T^2_{0,1} = \I \V'_{0,3}\,,\eqn\sabbythree$$

\refout

\bye